\documentclass[pra,aps,twocolumn,superscriptaddress]{revtex4-2}
\newcommand{\bra}[1]{\langle#1|}
\newcommand{\ket}[1]{\vert #1\rangle} 
\usepackage{xcolor}
\usepackage{epsfig}
\usepackage{amsmath}

\usepackage{ulem}

\newcommand{\mb}{\mathbf}
\newcommand{\expect}[1]{\langle#1\rangle}

\newcommand{\Pv}{$\mathcal{P}_V$}

\usepackage{amssymb}
\usepackage{amsfonts}
\usepackage{amsmath}
\usepackage{amsthm}

\usepackage{amsthm}
\usepackage{bm}

\begin{document}
	
\author{Mahasweta Pandit}
	\affiliation{Institute of Theoretical Physics and Astrophysics, Faculty of Mathematics, Physics and Informatics, University of Gdańsk, 80-308 Gdańsk, Poland}

	\author{Artur Barasi\'nski}
	\affiliation{
		Joint Laboratory of Optics of Palack\'y University and Institute of Physics of Czech Academy of Sciences, 771 46 Olomouc, Czech Republic}
	\affiliation{Faculty of Physics, University of Wroc{\l}aw, PL-50-204 Wroc{\l}aw, Poland}

\author{Istv\'an M\'arton}
	\affiliation{MTA Atomki Lend\"ulet Quantum Correlations Research Group,
		Institute for Nuclear Research, P.O. Box 51, H-4001 Debrecen, Hungary}

\author{Tam\'as V\'ertesi}
	\affiliation{MTA Atomki Lend\"ulet Quantum Correlations Research Group,
		Institute for Nuclear Research, P.O. Box 51, H-4001 Debrecen, Hungary}

	\author{Wies{\l}aw Laskowski}
	\affiliation{Institute of Theoretical Physics and Astrophysics, Faculty of Mathematics, Physics and Informatics, University of Gdańsk, 80-308 Gdańsk, Poland}
			\affiliation{International Centre for Theory of Quantum Technologies, University of Gdańsk, 80-308 Gdańsk, Poland}
	
	\title{Optimal tests of genuine multipartite nonlocality}

	\begin{abstract}
We propose an optimal numerical test for genuine multipartite nonlocality based on linear programming. In particular, we consider two non-equivalent models of local hidden variables, namely the Svetlichny and the no-signaling bilocal model. While our knowledge concerning these models is well established for Bell scenarios involving two measurement settings per party, the general case based on an arbitrary number of settings is a considerably more challenging task and very little work has been done in this field. 
In this paper, we applied such general tests to detect and characterize genuine $n$-way nonlocal correlations for various states of three qubits and qutrits. As a measure of nonlocality, we use the probability of violation of local realism under randomly sampled observables, and the strength of nonlocality, described by the resistance to white noise admixture. In particular, we analyze to what extent the Bell scenario involving two measurement settings can be used to determine genuine $n$-way non-local correlations generated for more general models. In addition, we propose a simple procedure to detect genuine multipartite nonlocality for randomly chosen settings with up to 100\% efficiency.
	\end{abstract}
	
	\maketitle

	\section{Introduction}

Genuine multipartite nonlocality (GMNL) is one of the most fundamental non-classical features of multipartite systems. Such genuinely $n$-way nonlocal correlations retained by multipartite entangled states are distinguishable from the correlations that are local to some bipartition. 
An inequality to test genuine tripartite nonlocality was first introduced by Svetlichny in Ref.~\cite{Svetlichny} and later generalized to multipartite scenarios~\cite{Seevinck,Collins}. Over time, these inequalities have been investigated and discussed extensively in many scientific papers (see e.~g.~\cite{Mitchell, Jones, Bancal}). Furthermore, in Ref.~\cite{Pironio}, two alternative definitions of genuine $n$-partite nonlocality were presented that are strictly weaker than that in Ref.~\cite{Svetlichny}, along with a series of Bell-type inequalities to detect genuine tripartite nonlocality. Such definitions are in line with an operational approach to GMNL, introduced independently in Ref.~\cite{Gallego}.

Genuinely $n$-way nonlocal correlations represent the most fragile form of genuine multipartite quantum correlations (GMQC) and are therefore at the top of the GMQC hierarchy.
Namely, for any $N$-partite qudit state, there exists a hierarchy \cite{Jia} such that genuine multipartite total correlations (GMT) $\supsetneq$ genuine multipartite discord (GMD) $\supsetneq$ genuine multipartite entanglement (GME) $\supsetneq$ genuine multipartite steering (GMS) $\supsetneq$ GMNL. Furthermore, various forms of GMQC are, in general, inequivalent to each other. For instance, it was shown that GMNL and GME are inequivalent for any number of parties \cite{Augusiak1}. In other words, there is a class of GME mixed states that are not GMNL \cite{Augusiak1, Augusiak2} and some of them are even fully local \cite{Bowles}. 
Exceptions are multipartite entangled pure states that are never fully local \cite{Popescu, Guhne} and for pure $n$-qubit symmetric states \cite{Chen} and all pure 3-qubit states \cite{Yu}, GME implies GMNL at the single-copy level. It was also shown that pure GME states can be GMNL if they have the structure of a network and also if there exist measurements that act on finitely many copies of any GME state to yield a GMNL behavior \cite{Tejada}. On the other hand, the GME mixed states that admit a fully local hidden variable model can be activated and display `hidden nonlocality’ if local operators can be applied prior to measurements \cite{Bowles,Gisin,Masanes1,Masanes2,Hirsch1,Hirsch2}. 

Recently, the violations of local realism by families of multipartite quantum states has been analysed numerically \cite{Laskowski} and experimentally \cite{Barasinskipra101_2020,
BarasinskiQuantum_2021,Jirakovaprapp16_2021}, showing that the probability of violation can also serve as a witness of GME. A notable advantage of these approaches is that the probability of violation operates in a so-called reference-frame-independent mode \cite{Chen2006PhysRevLett,Wabnig2013NJourPhys,Liu2019PhysRevApplied}, lowering requirements for alignment and calibration of remote devices \cite{Jirakovaprapp16_2021}. 

Despite the great effort put into the research of GMNL, these studies are mainly limited to the Bell scenario with two measurements per side. Consequently, the problem of the existence of GMNL and its characterisation in a general context (e.g with more than two measurements per side) remains open. 
In this work, we tackle these problems and use linear programming (LP) to analyze GMNL in the broader scenario. In this case, the only context information required is the number of parties, the number of measurements per party and the number of outcomes per measurement. We consider two measures of GMNL: the strength of nonlocality and the probability of violation of local realism. Since the complexity of the method increases rapidly with the number of parties, we focus on the case of three qubits and three qutrits. In Sec. \ref{method} we introduce the tools that are used to quantify nonlocality and derive LP constraints for the case of tripartite standard nonlocality and tripartite genuine nonlocality. In the case of genuine tripartite nonlocality, we consider Svetlichny-locality ($\mathcal{S}_2$) and no-signaling bilocality ($\mathcal{NS}_2$). In the next two sections, we present our numerical results and observations concerning the tripartite qubit and qutrit states, respectively. We also show that the three-qubit W state is $\mathcal{NS}_2$ nonlocal for four tetrahedrally distributed settings and three-qubit GHZ state is $\mathcal{S}_2$ nonlocal for two orthogonal planar settings.
	
	\section{The method}\label{method}
	
	We consider a Bell-type scenario in which three observers share a quantum state $\rho$, each of them can act locally on their shared part of the system. The parties are assumed to be spatially separated, and there is no communication between them during the course of the experiment. Each observer, labeled as $i = \{ 1,2,3 \}$, then have access to $m$ measurement choices $\{M_i^j\}_{j=1}^{m}$ that they can perform in their own distant laboratories. The correlations between measurement outcomes can be expressed in terms of the conditional probability distribution given by the expression 
	\begin{equation}
	p(r_1,r_2,r_3 | M_1^{i_1}, M_2^{i_2}, M_3^{i_3}) = {\rm Tr} (M_1^{i_1} \otimes M_2^{i_2} \otimes M_3^{i_3} \rho)  \label{prob},
	\end{equation}
	where, $\{r_{j}\}_{j=1}^{3}$ are the outcomes obtained by each party and $\{i_j\}_{j=1}^{3} \in [1,m]$.
	
	In our work, we consider two measures of genuine multipartite nonlocality: the strength of nonlocality and the probability of violation of local realism. 
	The strength of nonlocality $\mathcal{S}$ is related to the robustness to white noise. In other words, $\mathcal{S}$ can be quantified in terms of the amount of white noise that needs to be added to a convex mixture with $\rho$ in order to completely suppress the nonlocality of it, and the resulting state can be expressed as
	\begin{equation}
	\rho(v) = v \rho + (1-v) \rho_{\rm whitenoise},
	\end{equation}
	where $v$ is the visibility of the state. The strength of nonlocality can be defined as the smallest admixture of white noise for which the state becomes local for a fixed set of observables, $\mathcal{S} = 1 - v_{crit}$, where $v_{crit}$ is usually called critical visibility. Evidently, $\mathcal{S}$ can also be considered as a quantifier of nonlocality. 
	
	Finding the maximal $\mathcal{S}$ is a feasibility problem in linear programming. To determine $\mathcal{S}$, we consider the set of linear equations that is derived from relating the marginal probabilities to the underlying joint probability distributions for all possible results of measurements performed on $\rho(v)$. They have the following form
	\begin{widetext}
		\begin{equation}
		P(r_a, r_b,r_c|M_1^{a}, M_2^{b}, M_3^{c}) = \sum_{a_i,b_i,c_i = 1}^{d} p_{lr}(a_1,\dots,r_a,\dots,a_m, b_1,\dots,r_b,\dots,b_m,c_1,\dots,r_c,\dots,c_m,),
		\label{constrains}
		\end{equation}
	\end{widetext}
	where $\{a_i, b_i,c_i\}_{i=1}^{m}$ account for the hypothetical outcomes of the $i^{th}$ measurement, each performed by the individual parties and $d$ is the number of possible results. The set of linear constraints can be formed by considering marginal probabilities of the outcomes of all possible combinations of measurements that the three parties can independently perform in their laboratories. States that fail to satisfy (\ref{constrains}) do not attain local realism. $v_{crit}$ is obtained by increasing the value of  $v$ until the set of equalities (\ref{constrains}) can no longer be satisfied. In the next step, we optimize $v_{crit}$ over all possible measurement settings to determine the minimal critical visibility required for a given state which, in terms, gives us $\mathcal{S}^{max}= 1 - v_{crit}^{min}$.
	
	The second quantifier that we considered, denoted by $\mathcal{P}_{V}$, is the probability of violation of local realism. This quantity allows us to analyze the nonclassicality and correlation properties under random measurements \cite{Liang10}. $\mathcal{P}_{V}$ can be expressed in terms of a joint probability distribution when no specification on the Bell inequality is made. Such a probability distribution can be treated as an equivalent description of the analysis made by a full set of tight Bell inequalities in a given Bell scenario \cite{Barasinskipra99_2019}. $\mathcal{P}_{V}$ is then defined as
	\begin{equation}
	\mathcal{P}_{V}(\rho) = \int f(\rho, \Omega)~d\Omega,
	\end{equation}
	where the integration is over all parameters that vary within the Bell scenario and
	
	\begin{equation}
	f(\rho, \Omega) =
	\begin{cases}
	1, & \text{if settings lead to violations }\\
	& \text{of local realism,}\\
	0,& \text{otherwise.}
	\end{cases}       
	\end{equation}
	
	

	
\subsection{The formulation of LP constraints for tripartite standard and genuine nonlocality}
	
Below we introduce different notions of genuine multipartite Bell nonlocality beyond standard Bell nonlocality. In particular, we define the sets $\mathcal{L}$, $\mathcal{S}_2$ and $\mathcal{NS}_2$ for three parties. We follow the works of Refs.~\cite{GTE1,GTE2}. We show that in order to decide if a given correlation point $\mb{p}$ is within one of the above sets or lies outside the set is a linear programming problem. We provide the LP task to be solved and we also look at the problem from a geometrical point of view. 

{\it Introducing the definitions of the sets}---We define the sets $\mathcal{L}$, $\mathcal{S}_2$ and $\mathcal{NS}_2$. We consider three separated parties, Alice, Bob and Charles, who can conduct $m$ different measurements with two outcomes ($o=2$) each. We use the notation $a=r_1$, $b=r_2$, $c=r_3$ and $x=i_1$, $y=i_2$, $z=i_3$ in Eq.~(\ref{prob}). Namely, we denote the measurement settings by $x,y,z$ and their outcomes by $a,b,c$. The number of inputs are $m$, that is, $x=(1,2,\ldots,m)$, $y=(1,2,\ldots,m)$ and $z=(1,2\ldots,m)$ and their measurement outcomes are $a=(0,1)$, $b=(0,1)$ and $c=(0,1)$. Then a three-party Bell experiment is characterized by the joint probability distribution $P(a,b,c|x,y,z)$ consisting of $8\times m^3$ values. If the probability distribution can be written as
\begin{equation}
P(a,b,c|x,y,z)=\sum_{\lambda}q_{\lambda}P_\lambda(a|x)P_{\lambda}(b|y)P_{\lambda}(c|z),
\label{Pstandard}
\end{equation}
with $0\le q_\lambda\le 1$ and $\sum_{\lambda}q_{\lambda}=1$, then the correlation point $\mb{p}=\{P(a,b,c|x,y,z)\}$ is inside the set $\mathcal{L}$, and we say that the point $\mb{p}$ is Bell local or Bell classical. Otherwise, we say that the correlation $\mb{p}$ is Bell nonlocal or nonclassical.  

As pointed out by Svetlichny~\cite{Svetlichny}, some correlations $\mb{p}\not\in\mathcal{L}$ can be still written in the hybrid local-nonlocal form:
\begin{align}
P(a,b,c|x,y,z)&=\sum_{\lambda_1}q_{\lambda_1}P_{\lambda_1}(a,b|x,y)P_{\lambda_1}(c|z)\nonumber\\
&+\sum_{\lambda_2}q_{\lambda_2}P_{\lambda_2}(a,c|x,z)P_{\lambda_2}(b|y)\nonumber\\
&+\sum_{\lambda_3}q_{\lambda_3}P_{\lambda_3}(b,c|y,z)P_{\lambda_3}(a|x),
\label{Psvet}
\end{align}
where $0\le q_{\lambda_i}\le 1$ for $i=1,2,3$ and $\sum_{\lambda_1} q_{\lambda_1}+\sum_{\lambda_2} q_{\lambda_2}+\sum_{\lambda_3} q_{\lambda_3}=1$. Note that above there is no restriction on the form of the bipartite probability distributions $P_{\lambda_1}(a,b|x,y)$, $P_{\lambda_2}(a,c|x,z)$ and $P_{\lambda_3}(b,c|y,z)$. In particular, they can allow signaling as well. We call the correlation $\mb{p}$ written in the form of Eq.~(\ref{Psvet}) as Svetlichny-local, or $S_2$ correlations for brevity. These correlations are inside the set $\mathcal{S}_2$.

On the other hand, following the definition of Ref.~\cite{GTE1}, a tripartite correlation with probabilities $P(a,b,c|x,y,z)$ is no-signaling bilocal ($NS_2$) and is inside the set $\mathcal{NS}_2$ if it can be written as
\begin{align}
P(a,b,c|x,y,z)&=\sum_{\lambda_1}q_{\lambda_1}P_{\lambda_1}^{\text{NS}}(a,b|x,y)P_{\lambda_1}(c|z)\nonumber\\
&+\sum_{\lambda_2}q_{\lambda_2}P_{\lambda_2}^{\text{NS}}(a,c|x,z)P_{\lambda_2}(b|y)\nonumber\\
&+\sum_{\lambda_3}q_{\lambda_3}P_{\lambda_3}^{\text{NS}}(b,c|y,z)P_{\lambda_3}(a|x),
\label{PNS2}
\end{align}
where the bipartite correlations (for instance, $P_{\lambda_1}^{\text{NS}}(a,b|x,y)$) respect the no-signaling constraints, which are given by 
\begin{align}
\sum_a P_{\lambda_1}^{\text{NS}}(a,b|x,y)&=\sum_a P_{\lambda_1}^{\text{NS}}(a,b|x',y)\text{ for all } x,x',y,b,\lambda_1\nonumber\\
 \sum_b P_{\lambda_1}^{\text{NS}}(a,b|x,y)&=\sum_b P_{\lambda_1}^{\text{NS}}(a,b|x,y')\text{ for all } y,y',x,a,\lambda_1,
\end{align}
and similar relations obtained from permutations of the parties. If at least one of the above constraints is not satisfied, then this allows for signaling. All the no-signaling constraints are linear in the function of probabilities. Note that we have the inclusion relations $\mathcal{L}\subsetneq\mathcal{NS}_2\subsetneq\mathcal{S}_2$ between the above defined sets~\cite{GTE1}.

{\it Formulating the membership problem as an LP}---Given a correlation point $\bm{p}$, we now give a linear programming algorithm for detecting the point $\mb{p}$ inside the sets $\mathcal{L}$, $\mathcal{NS}_2$ or $\mathcal{S}_2$.
The first step is to realize that any randomness present in either the two-party or one-party response functions $P_{\lambda}$, $P_{\lambda_i}$, $P_{\lambda_i}^{\text{NS}}$ for $i=1,2,3$ in Eqs.~(\ref{Pstandard},\ref{Psvet},\ref{PNS2}) can always be incorporated in the shared random variable $\lambda$ and $\lambda_i$, $i=1,2,3$. Hence we can assume that these functions are deterministic functions of the inputs $x$, $y$ and $z$. In other worlds, each shared random variable $\lambda$, $\lambda_i$, $i=1,2,3$ defines an assigment of one of the possible outputs to each input. There can only be a finite number of such assignments and we name them as $D_{\lambda}(a|x)$, $D_{\lambda_i}(a,b|x,y)$, $D_{\lambda_i}^{\text{NS}}{(a,b|x,y)}$ for $i=1,2,3$ and similarly for the other two possible $(AC|B)$ and $(BC|A)$ bipartitions. Therefore, we can assume without the loss of generality that in Eqs.~(\ref{Pstandard},\ref{Psvet},\ref{PNS2}) the functions are the above deterministic functions. For instance,  equation~(\ref{Pstandard}) can be equivalently written as
\begin{equation}
P(a,b,c|x,y,z)=\sum_{\lambda}q_{\lambda}D_\lambda(a|x)D_{\lambda}(b|y)D_{\lambda}(c|z).
\label{Dstandard}
\end{equation}
Let us then denote $\mb{d}_\lambda$ such a product of these deterministic functions. E.g. in equation~(\ref{Dstandard}), we have $\mb{d}_{\lambda}=D_\lambda(a|x)D_{\lambda}(b|y)D_{\lambda}(c|z)$. Then a correlation $\mb{p}$ is local (within $\mathcal{L}$) if it can be written as a convex combination of deterministic functions $\mb{d}_{\lambda}$, that is
\begin{equation}
\mb{p}=\sum_\lambda q_\lambda \mb{d}_\lambda,\text{ with } q_\lambda \geq 0,\, \sum_\lambda q_\lambda=1\,.
\label{Lfeas}
\end{equation}
Similarly, in the $S_2$ and $NS_2$ cases we have 
\begin{equation}
\mb{p}=\sum_{\lambda_1} q_{\lambda_1} \mb{d}_{\lambda_1}+\sum_{\lambda_2} q_{\lambda_2} \mb{d}_{\lambda_2}+\sum_{\lambda_3} q_{\lambda_3} \mb{d}_{\lambda_3}
\label{NS2feas}
\end{equation}
with
\begin{equation}
q_{\lambda_1}, q_{\lambda_2}, q_{\lambda_3} \geq 0,\, \sum_\lambda q_{\lambda_1}+\sum_\lambda q_{\lambda_2}+\sum_\lambda q_{\lambda_3}=1\,.
\end{equation}
These representations are very useful as they provide an algorithm for determining if a given correlation $\mb{p}$ is within the above sets $\mathcal{L}$, $\mathcal{S}_2$ or $\mathcal{NS}_2$~\cite{Zuk99,Kas00}. 

Indeed, determining whether there exist weights $\lambda$ in equation~(\ref{Lfeas}) and weights $\lambda_i$, $i=1,2,3$ in equation~(\ref{NS2feas}) satisfying the given linear constraints is a linear programming problem, for which there exist efficient solvers (i.e. which run in polynomial time in the number of variables~\cite{LP}). However, there are $2^{3m}$ possible $\lambda$ values in the $L$ problem~(\ref{Lfeas}), and there are $12\times 2^m m^2$ in the $S_2$ problem~(\ref{NS2feas}).  In fact for these cases (and for $NS_2$ as well) the scaling of the number of $\lambda$ values is exponential in $m$, hence the algorithm is not efficient by itself.

Now we turn (\ref{Lfeas}) and (\ref{NS2feas}) feasibility problems into a problem of finding the critical visibility of a probability distribution $\mb{p}$ such that the noisy distribution 
\begin{equation}
\mb{p}_v=v\mb{p}+(1-v)\mb{p}_{\text{iso}}
\label{pv}
\end{equation}
can be still described within the respective sets of 
$\mathcal{L}$, $\mathcal{S}_2$ and $\mathcal{NS}_2$, where $\mb{p}_{\text{iso}}$ is the 
isotropic distribution defined by $P_{\text{iso}}(a,b,c|x,y,z)=1/2^3$ for all $a,b,c$ and $x,y,z$. Then finding the critical visibility in the membership problem for $\mathcal{L}$ translates to the following LP task:
\begin{align}
v_{crit} &= \max v\nonumber\\
\text{ s. t. } v\mb{p}+
(1-v)\mb{p}_{\text{iso}} &= \sum_{q_{\lambda}} q_{\lambda}\mb{d}_{\lambda}\nonumber\\
\text{with } q_{\lambda}&\geq  0,\, \sum_\lambda q_{\lambda}=1\,.
\label{LPvcritL}
\end{align} 
On the other hand, in the case of $S_2$ and $NS_2$ problems we compute the critical visibility via LP as follows:
\begin{align}
v_{crit} &= \max v\nonumber\\
\text{ s. t. } v\mb{p}+
(1-v)\mb{p}_{\text{iso}} &= \sum_{i=1,2,3}\sum_{q_{\lambda_i}} q_{\lambda_i}\mb{d}_{\lambda_i}\nonumber\\
\text{with } q_{\lambda_i}&\geq  0,\, \sum_{i=1,2,3}\sum_\lambda q_{\lambda_i}=1\,.
\label{LPvcritS2}
\end{align} 

Since each of the above sets $\mathcal{L}$, $\mathcal{S}_2$ and $\mathcal{NS}_2$ is the convex hull of a finite number of points, each set defines a polytope. The local deterministic points $\mb{d}_{\lambda}$ correspond to the vertices, or extreme points, of the polytope. Then we can use the dual formulation of the above problems~(\ref{LPvcritL}) and (\ref{LPvcritS2}) to get a hyperplane (defined by the normal plane $\mb{S}$ and a constant $S$), which separates the point $\mb{p}$ from the respective polytope.  This way by solving the dual LP we obtain a linear inequality $\mb{S}\cdot\mb{d}_{\lambda}\le S$ satisfied by all the vertices $\mb{d}_{\lambda}$, but violated by the correlation point $\mb{p}(v)$: $\mb{S}\cdot\mb{p}(v)\ge S$ for $v>v_{crit}$.

The above formalism can be generalized to a larger number of parties. For example, in the case of the four-party Svetlichny set, we obtain a decomposition similar to the three-party decomposition~(\ref{Psvet}), but in addition to the partitions of type $(A|BCD)$ (involving single-party terms), there are partitions of type $(AB|CD)$ (not involving single-party terms). That is, we will have factorizations of probabilities with different types of cuts, such as $P_{\lambda_i}(a,b|x,y)P_{\lambda_i}(c,d|z,w)$ and $P_{\lambda_{i'}}(a,b,c|x,y,z)P_{\lambda_{i'}}(d|w)$. However, the problem is still linear and involves a finite number of different possible $\lambda$ variables, and should not pose a problem for solving the LP, apart from the fact that the complexity of the optimization problem increases rapidly with the number of parties.

\section{Results -- qubits}
	
We now apply our optimization method to different tripartite states, which are often considered as nonclassical crucial resources:
\begin{itemize}
\item[$-$] the Greenberger - Horne - Zeilinger (GHZ) state \cite{Greenberger1989, Scarani_2001}
\begin{equation}
\ket{{\rm GHZ}}=\frac{1}{\sqrt{2}}( \ket{000}+\ket{111}) \label{ghz}
\end{equation}
\item[$-$] the W state \cite{PhysRevA.62.062314}
\begin{equation}
\ket{{\rm W}}=\frac{1}{\sqrt{3}}( \ket{100}+\ket{010}+\ket{001}),    
\end{equation}
\item[$-$] the $\psi_s$ state mentioned in Ref.~\cite{PhysRevA.94.062121} that is equally entangled in every bipartition
\begin{equation}
\ket{{\psi_{s}}}= \frac{1}{\sqrt{6}}(|001\rangle + |010\rangle - |100\rangle) + \frac{1}{\sqrt{2}} |111\rangle.    
\end{equation}
\end{itemize}

		\begin{table*}[ht]
	\centering
	\begin{tabular}{ cc|cc|cc|cc|cc|cc|cc }
 \hline \hline
 &  &  \multicolumn{4}{c}{Standard nonlocality} & \multicolumn{8}{|c}{Genuine multipartite nonlocality}\\\hline
		State   & $m$ & \multicolumn{4}{c}{$\mathcal{L}$} & \multicolumn{4}{|c}{$\mathcal{S}_2$} & \multicolumn{4}{|c}{$\mathcal{NS}_2$}  \\\hline
			& & $\mathcal{P}_{V}$ & $\mathcal{S}_{max}$  & $\mathcal{P}_{V}^{I}$ & $\mathcal{S}_{max}^{I}$ & $\mathcal{P}_{V}$ & $\mathcal{S}_{max}$ & $\mathcal{P}_{V}^{I}$ & $\mathcal{S}_{max}^{I}$  & $\mathcal{P}_{V}$ & $\mathcal{S}_{max}$ & $\mathcal{P}_{V}^{I}$ & $\mathcal{S}_{max}^{I}$  \\ \hline
			 GHZ & 2 & 74.69 \cite{Laskowski} & 0.5 & 70.00 & 0.2929 			& 0.5413 & 0.292893 	& 0.5353 & 0.292893 & 11.57 &  0.292893 & 10.6270 &  0.1716 \\
		    	    & 3 & 99.54\cite{Laskowski} & 0.5 & 99.22 &  				& 6.6282 & 0.292893 	& 6.6009 &  		& 70.81 &  0.292893 & 67.2124 &   \\
		 	   & 4 & $>99.99$ \cite{Laskowski}& 0.50301 \cite{PhysRevA.84.042122} & $>99.99$ &  		& 23.255 & 0.292893 	& 22.9506 &  		& 96.73 & 0.292893 	& 95.6705 &   \\
		   	   & 5 & $>99.99$ \cite{Laskowski}& 0.50481  & $>99.99$ &  			& 46.01 & 0.292893    	& 45.5509 &  		& 99.77 & 0.292893	& 99.6375 &   \\
		   	   & 6 & $>99.99$ \cite{Laskowski}& 0.50639 & $>99.99$ &  			& 66.67 & 0.292893	  	& 66.6914 &  		& 99.99 & 0.292893	& 99.9731 &   \\ \hline
	 	   W   & 2 & 54.89 \cite{Laskowski}& 0.3558 \cite{PhysRevA.82.012118} & 50.86  & 0.3558 		& 0.0085 &0.082107 		& 0.0030 & 0.08144 	& 3.73 	& 0.199118 	& 3.2312  &  0.1895 \\
			      & 3 & 97.80 \cite{Laskowski} & 0.3950 \cite{PhysRevA.82.012118} & 96.70  &  			& 0.01951 & 0.082107 	& 0.0696 &  		& 41.29 & 0.199198 	& 37.2382  &   \\
			     & 4 & $>99.99$ \cite{Laskowski} & 0.3985\cite{PhysRevA.82.012118} & 99.92 &  			& 1.3797 & 0.082107 	& 0.4679 &  		& 86.69 & 0.199860 	& 83.1669  &   \\
		  	      & 5 & $>99.99$ \cite{Laskowski}& 0.4009 & $>99.99$  &  			& 5.49 & 0.082107		& 1.7823 &  		& 98.71 & 0.203114	& 98.0032  &   \\
		  	      & 6 & $>99.99$ \cite{Laskowski}& 0.4044 & $>99.99$  &  			& 14.33 &  0.082107		& 4.8795 &  		& 99.93 & 0.204638	& 99.8638  &   \\ \hline
$\ket{{\psi_{s}}}$ & 2 & 64.38 & 0.46445 & 64.38 & 0.30094	& 0.2019 & 0.242821		& 0.2055 & 0.242821 & 6.8637 & 0.242821	& 5.4267 &  0.1700 \\ 
                     & 3 & 99.03 & 0.46455 & 98.21 &  		& 2.9871 & 0.242821		& 2.9877 &  		& 58.9212 & 0.242821& 49.9979 &   \\
                     & 4 & $>99.99$ & 0.4669 & 99.97 &  		& 12.3838 & 0.242821	& 12.3200 &  		& 94.55 & 0.247833	& 89.9767 &   \\
                     & 5 & $>99.99$ & 0.46699 & $>99.99$ & 	& 28.91 & 0.242821		& 28.6249 &  		& 99.6594 & 0.247856& 98.9490 &   \\
                     & 6 & $>99.99$ &0.46747 & $>99.99$ &  	& 48.32 & 0.242821		& 47.9802 &  		& $>99.99$ & 0.247856			& 99.9158 &   \\
			\hline \hline
		\end{tabular}
		\caption{\label{tab_qubits} Numerically obtained $\mathcal{P}_{V}$ (in \%) and $\mathcal{S}_{max}$ for an increasing number of measurement settings $m$ per site and different nonlocality scenarios: standard nonlocality: $\mathcal{L}$; genuine multipartite nonlocality: $\mathcal{S}_2$, $\mathcal{NS}_2$. $P_V^I$ denotes the probability of violation (in \%) for 'dominant Bell inequality'.  In the same way  $\mathcal{S}_{max}^{I}$ is defined.}
	\end{table*}

We investigate probabilities of violation $\mathcal{P}_{V}$ and the maximal nonlocality strengths $\mathcal{S}_{max}$ for these states with an increasing number of measurement settings per site and for $\mathcal{L}$, $\mathcal{S}_2$, $\mathcal{NS}_2$ scenarios (see Tab \ref{tab_qubits}). The following observations are drawn:

(1) Obviously, $\mathcal{P}_{V}$ steadily increases with $m$. 
The $\mathcal{L}$ condition is violated almost with certainty once the number of settings $m > 2$ for the states that has been considered and we see a probability of violation greater than $99.99\%$ for $m > 3$. 
Similar, although slightly weaker, behavior is observed concerning the $\mathcal{NS}_2$ constraint violation. 
While for $m=2$, $\mathcal{P}_v$ values are significantly smaller than that of the $\mathcal{L}$ case, the difference decreases for $m \geq 4$ for the GHZ state and $m \geq 5$ for the W and $\psi_s$. However, in the case of $\mathcal{S}_2$, the percentage is considerably smaller, especially for the W state, even for $m=6$.

A natural question arises about the nature of the increase in $\mathcal{P}_{V}$ with an increase in the number of settings $m$. This property might arise either by the growing number of equivalent Bell inequalities (compared to the $m=2$ case) or by the emergence of new tight Bell inequalities. 
To answer this question, we consider families of tight Bell inequalities that represent a complete set of facets characterizing the $\mathcal{L}$ and $\mathcal{NS}_2$ polytopes (with $m=2$), provided in Ref. \cite{Pitowskypra64_2001,Sliwapla317_2003} and \cite{GTE1}, respectively. As the complete characteristic of the $\mathcal{S}_2$ polytope in terms of Bell inequalities remains unknown, in this case, we restrict our considerations to the symmetric facets discussed in Ref.~\cite{Bancal_2010}. 

Our calculations provide that in all scenarios discussed above one can clearly distinguish a so-called dominant Bell inequality, i.e. the family of equivalent Bell inequalities, which are most often violated for a given Bell experiment \cite{Barasinskipra99_2019,BarasinskiQuantum_2021}. For the the GHZ and $\ket{{\psi_{s}}}$ states such dominant position is reserved for the 4th facet inequality ($I_4^{\mathcal{L}}$) for the $\mathcal{L}$ scenarios, the 4th facet inequality ($I_4^{\mathcal{NS}_2}$) for the $\mathcal{NS}_2$ scenarios, and Svetlichny’s original inequality ($I_{185}^{\mathcal{S}_2}$) for the $\mathcal{S}_2$ polytope. For the W state one should additionally consider the 16th facet inequality ($I_{16}^{\mathcal{NS}_2}$) of the $NS_{2}$ polytope.  
Note that all inequalities are numbered in the same manner as in Ref. \cite{Sliwapla317_2003} and \cite{GTE1}.

As we see in Tab. \ref{tab_qubits}, despite the great limitations of the problem \cite{Barasinskipra99_2019,BarasinskiQuantum_2021}, the estimated probability of violation (hereafter denoted as $\mathcal{P}_{V}^{I}$) is surprisingly consistent with the previous calculations for almost all Bell experiments. In particular, a very good consistency is observed for the $\mathcal{L}$ and $NS_{2}$ scenarios. The worst result has been found for the W states with $\mathcal{S}_2$ constraints, despite a very good approximation for other scenarios. It suggests that either the Svetlichny inequality is not the dominant inequality for W states or there is no dominant inequality for this case at all. Since we know only symmetric facets of the $\mathcal{S}_2$ polytope, we cannot answer this question.

Next, using the lifting procedure \cite{Pironiojmp46_2005}, all dominant Bell inequalities have been adapted to the $m>2$ Bell scenario. Similarly to the previous results, the probability of violation $\mathcal{P}_{V}^{I}$ increases with increasing number of settings $m$. Moreover, except for the W states and $\mathcal{S}_2$ constraint, the gap between $\mathcal{P}_{V}^{I}$ and the linear-programming results vanishes with growing $m$. It implies that for these cases the increase of $\mathcal{P}_{V}$ with $m$ has a statistical explanation, i.e. by increasing the number of settings, we increase the number of equivalent Bell inequalities that belong to the dominant family and hence, the probability that some of them are violated, involving only two settings. 

(2) For W state violating the $\mathcal{L}$ constrains, $\mathcal{S}_{max}$ practically appears to be dependent on the number of settings. For other cases, however, either we do not observe any dependence up to 6 decimal places (GHZ: $\mathcal{S}_2$, $\mathcal{NS}_2$) or we see slight improvement on the 4th of 5th decimal place (GHZ: $\mathcal{L}$; W: $\mathcal{S}_2$, $\mathcal{NS}_2$; $\psi_2$: $\mathcal{L}$). Note that the above improvements are a real effect and not the result of numerical inaccuracy. Comparing these results with the estimates obtained for the dominant Bell inequalities ($\mathcal{S}_{max}^{I}$), we can see that such inequalities usually do not provide a correct evaluation of $\mathcal{S}_{max}$ even for $m = 2$. Moreover, the improvement of the $\mathcal{S}_{max}$ for $m>2$ implies the emergence of new types of Bell inequalities. 
In other words, $\mathcal{S}_{max}$ solely depends on the chosen Bell experiment and, in general, cannot be explained by the scenario involving only two settings. The exception is the $\mathcal{S}_2$ scenario, in which, regardless of the chosen $m$, $\mathcal{S}_{max}$ is provided either by the Svetlichny inequality (GHZ state, $\ket{{\psi_{s}}}$ state) or by the symmetric inequality no. 31 derived in Ref. \cite{Bancal_2010} (W state).

\begin{table*}[ht]
\begin{tabular}{c ccccc|ccccc} \hline \hline
&\multicolumn{5}{c}{$\mathcal{P}_{V}$} & \multicolumn{5}{|c}{$\mathcal{P}_{V}^{I}$} \\ \hline
$m$ &  $10^o$ &  $20^o$ &  $30^o$  & $40^o$ & $45^o$  &  $10^o$ &  $20^o$ &  $30^o$  & $40^o$ & $45^o$ \\ \hline
2  & 0.97 &      4.13 & 8.52  &  11.27 &     11.57 
& 0.29 &      2.92 & 7.52  &  10.35 &     10.63 \\
3  & 13.64 &      41.62 & 63.61  &  70.52 &     70.81 
& 3.96 &      29.57 & 58.02  &  67.02 &     67.21\\
4  & 43.83 &      84.73 & 95.81  &  \textbf{96.84} &     \textbf{96.72} 
& 14.37 &      68.45 & 93.63  &  \textbf{95.89} &     \textbf{95.67}\\
5  & 72.49 &      98.07 & \textbf{99.79}  &  \textbf{99.80} & \textbf{99.77} 
& 29.83 &      90.46 & 99.59  &  \textbf{99.70} & \textbf{99.64}\\
6  & 88.29 &      99.85 & \textbf{99.993} &  \textbf{99.990} &   \textbf{99.987} 
& 46.48 &      97.80 & \textbf{99.984} &  \textbf{99.982} &   \textbf{99.973}\\ 
\hline \hline
\end{tabular}
\caption{\label{tab:results2} Probability of violation (in \%) of $\mathcal{NS}_2$ model for the asymmetric GHZ state, $|GHZ(\alpha)\rangle = \cos \alpha |000\rangle + \sin \alpha |111\rangle$, determined from $10^7$ ($m=2$) and $10^6$ ($m>2$) random samples and various angles $\alpha$.}
\end{table*}
	\vspace{1cm}
	
(3) We further investigate $\mathcal{P}_{V}$ for violating $\mathcal{NS}_2$ constraints using the asymmetric GHZ state given by $|GHZ(\alpha)\rangle = \cos \alpha |000\rangle + \sin \alpha |111\rangle$ (see Tab.~\ref{tab:results2}). $\mathcal{P}_{V}$ exhibits interesting effect with respect to the increasing value of $\alpha$. For $m=2,3,4$ we observe expected monotonic increase of \Pv~with $\alpha$. However, this monotonicity is disrupted for $m=4,5$ where there is a local decrease of \Pv (minimum) around $\alpha=40^o$.  Although the effect manifests itself at the 4th decimal place, it is very likely not due to numerical inaccuracies (an analysis based on confidence interval is given below). It is worth noting that a similar effect is not observed for the standard nonlocality problem.

Interestingly, a similar phenomenon can be observed if the 4th facet inequality of $\mathcal{NS}_2$ \cite{GTE1} is taken into consideration. As presented in Table~\ref{tab:results2}, the monotonicity of \Pv ~for $I_4^{\mathcal{NS}_2}$ is disrupted for $m>4$ even more than for the entire polytope  $\mathcal{NS}_2$. Our analysis suggests that this effect may be caused by the overlapping of the parameter space leading to a violation of a single inequality $I_4^{\mathcal{NS}_2}$. To be clear, let $I_4^{\mathcal{NS}_2}$ be the original Bell inequality given in Ref. \cite{GTE1} and $\mathcal{I}_4^{\mathcal{NS}_2}$ its equivalent form obtained by permutation of inputs/outputs/parties. Furthermore, let us define a parameters subspace $\omega$ ($\gamma$) which contains all settings leading to violation of $I_4^{\mathcal{NS}_2}$ ($\mathcal{I}_4^{\mathcal{NS}_2}$). One can easily show that the intersection $\omega \cap \gamma \neq \emptyset$ for any $\mathcal{I}_4^{\mathcal{NS}_2}$.  Therefore, the overall $\mathcal{P}_{V}^{I}$ cannot be considered as the probability of violation for a single Bell inequality (the cardinality of $\omega$) multiplied by the number of equivalent inequalities and the degree of intersection, which depends on the state under consideration, has to be taken into consideration. For that reason, although the probability of violation for a single Bell inequality $I_4^{\mathcal{NS}_2}$ reveals the monotonicity with respect to $\alpha$, such property may be disrupted for the total $\mathcal{P}_{V}^{I}$ if $m$ is sufficiently high.

To support the above claims, we analyze the problem statistically by interpreting the probability of violation ($\mathcal{P}_v$) as the success probability of a Bernoulli trial~\cite{Papoulis2002}. In a Bernoulli trial, one performs independent repeated trials of a random experiment with two possible outcomes. One outcome is called ``success'' and the other outcome is called ``failure''. In our case ``success'' is when a quantum correlation point is genuinely tripartite nonlocal. For a similar interpretation of the relative volume for the set of bipartite quantum correlations, see the recent paper~\cite{Lin2021}. 

In a Bernoulli trial process, $p = n_s/n$ is the proportion of success, measured by $n$ trials yielding $n_s$ successes. For a confidence level of $99\%$, the error is $\epsilon = 0.01$ and $z = 2.576$ (which is the quantile $(1-\epsilon/2)$ of the standard normal distribution). From $z$ and $p$, we can estimate the value of $\mathcal{P}_v$, giving lower and upper bounds on it based on the central limit theorem. However, this approximation is not reliable if the sample size is small or if $\mathcal{P}_v$ is close to 0 or 1. In our case, we would like to compare two values of $\mathcal{P}_v$ corresponding to $m=4$ settings and angles of $\alpha = 40^o$ and $\alpha = 45^o$. The respective proportion of successes are $p_1=0.9684$ and $p_2=0.9672$ (see Tab.~\ref{tab:results2}). Since these values are relatively close to 1, we use the Wilson score interval instead of the normal approximation interval for the estimation of $\mathcal{P}_v$. This interval can be calculated directly from Wilson's formula for a given confidence level~\cite{Wilson1927}. This gives the following estimates of the probability of violation for a confidence level of $99\%$: 
\begin{align}
&0.9679<\mathcal{P}_v^1<0.9688,\nonumber\\ 
&0.9667<\mathcal{P}_v^2<0.9677.
\end{align}
As can be observed, there is no overlap between the two above intervals. Thus, with a confidence level of $99\%$ we can state that the probability of violation $\mathcal{P}_v$ is indeed slightly higher around the angle $\alpha=40^o$ than around the angle $\alpha=45^o$. The same conclusion can be drawn for $\mathcal{P}_v^I$ for $m=4$ (see Tab.~\ref{tab:results2}). In this case we can also conclude with a confidence level of $99\%$ that $\mathcal{P}_v^I$ is larger for the angle $\alpha = 40^o$ than for the angle $\alpha = 45^o$.

	\section{Results -- qutrits}
	
We now investigate  $\mathcal{P}_{V}$ and $\mathcal{S}_{max}$ for qutrits but limiting it to $\mathcal{L}$ and $\mathcal{NS}_2$-type violation. We run our investigation for $m=2$ and $m=3$ measurement settings considering a range of three qutrit states given by:
	
\begin{itemize}
   	   \item[$-$] the three qutrit GHZ-type state 
	\begin{equation}
	    \ket{{\rm GHZ}}=\frac{1}{\sqrt{3}}(\ket{000}+\ket{111}+\ket{222}),
	\end{equation}
	   \item[$-$] the three qutrit GHZ-type state reduced to qubit subspace (rank-2 state)
	\begin{equation}
	    \ket{{\rm rank-2}} = \frac{1}{\sqrt{2}}(\ket{000}+\ket{111}),
	\end{equation}
	   \item[$-$] the three qutrit Dicke states \cite{Laskowski_2014}
	\begin{eqnarray}
	    \ket{D_{3}^{1}}&=& \frac{1}{\sqrt{3}}(\ket{001} + \ket{010} + \ket{100}), \\
		\ket{D_{3}^{2}}&=& \frac{1}{\sqrt{15}} (\ket{002} + \ket{020} + \ket{200} \\ 
		&+& 2 (\ket{011} + \ket{101} + \ket{110})), \nonumber\\
		\ket{D_{3}^{3}}&=& \frac{1}{\sqrt{10}}(\ket{012} + \ket{021} + \ket{102}  \\ &+& \ket{120} + \ket{201} + \ket{210} + 2 \ket{111}), \nonumber
	\end{eqnarray}
	  	   \item[$-$] the three qutrit singlet state \cite{PhysRevLett.87.217901, PhysRevLett.89.100402}
	\begin{eqnarray}
		    \ket{\rm{Aharonov}}&=&\frac{1}{\sqrt{6}}(\ket{012} - \ket{021} -\ket{102}\\ &+&\ket{120} +\ket{201} - \ket{210}). \nonumber
	\end{eqnarray}
	\end{itemize}

The numerical results are provided in Tab. \ref{tab_qutrits}. At first glance, it is striking that the probability of witnessing genuine multipartite nonlocality ($\mathcal{NS}_2$) is negligible, and is only noticeable (16.56\%) in the case of the the GHZ state for measurement settings $m=3$. If we compare these probabilities of violation (for $m=2$) with the values for the case of standard nonlocality ($\mathcal{L}$), it turns out that they are smaller by two (GHZ, rank-2), three ($D_3^1$, $D_3^2$, $D_3^3$) and even five (Aharonov) orders of magnitude. 
	
However, the  probability of violation for the $\mathcal{NS}_2$  problem increases up to 17-30 times depending on the state with measurement settings set to $m=3$. For qubits, the increment is smaller and varies in the range of 6-11 times. Of course, it is difficult to compare these two cases due to the fact that different states were considered in both cases. If we limit ourselves to comparing only the family of the GHZ states then in the case of qubits the increment was 6.12 times, while for qutrits it is 30.02 times. 

	\begin{table}
		\begin{tabular}{cccc} \hline \hline
State &  $m$ &  $\mathcal{L}$  & $\mathcal{NS}_2$   \\ \hline
GHZ         &2& 81.1435 & 0.5516 \\
            &3& 99.9996 & 16.5618 \\
rank-2      &2& 33.1366 & 0.1854 \\
            &3& 92.2090 & 3.2805 \\
$D_3^1$     &2& 17.2009 & 0.0190 \\
            &3& 73.4908 & 0.3483 \\
$D_3^2$     &2& 39.3363 & 0.0506\\
            &3& 98.5799 & 1.2658\\
$D_3^3$     &2& 42.5216 & 0.0807\\
            &3& 99.6149 & 2.0036\\
Aharonov    &2& 68.7038 & 0.0056 \\ 
            &3& 99.9859 & 0.1643 \\ 
				\hline \hline	
		\end{tabular}
		\caption{\label{tab_qutrits} Probabilities of violation (in $\%$) obtained for qutrit states from $10^6$ random samples for $\mathcal{L}$ and $\mathcal{NS}_2$ models.}
	\end{table}

	\section{Three-qubit W state is $\mathcal{NS}_2$ nonlocal for four tetrahedrally distributed  settings}\label{qubitw}
	
\begin{figure}[ht]
		\centering
		\includegraphics[width=0.45 \textwidth]{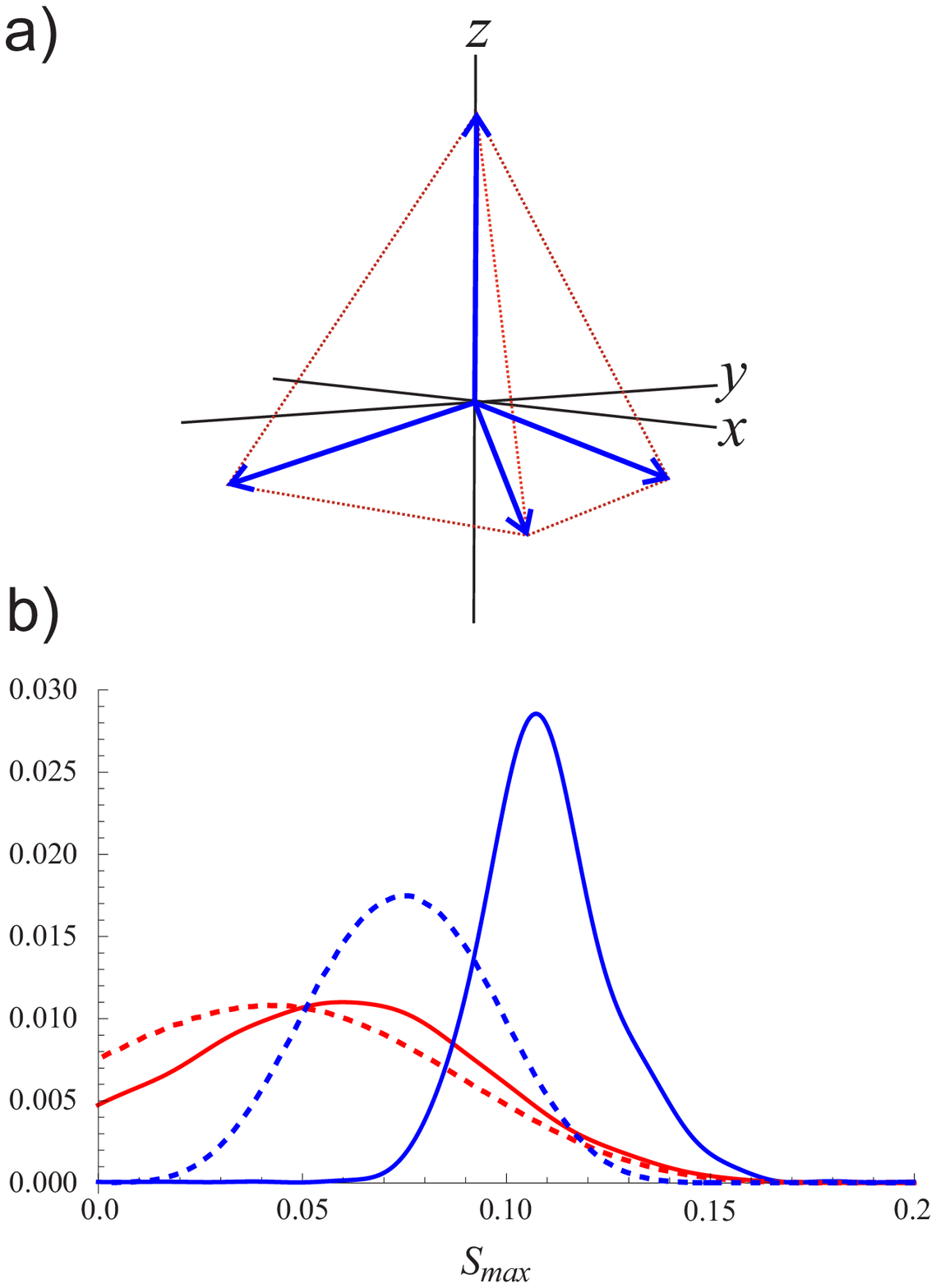}
		\caption{\label{fig:tetrads} (a) Tetrahedral distribution of measurement settings; (b) Probability density function for $\mathcal{P}_{V}$ (solid) and $\mathcal{P}_v^I$ (dashed) for the W state: tetrahedrally (red) vs. uniformly distributed settings (blue).}
			\end{figure}

	We consider an experimental situation in which three observers can choose between four seetings forming a regular tetrahedron (see Fig. \ref{fig:tetrads}(a)). The corresponding observables are given by: 
 $O_j^i = U^i o_j^i (U^i)^{\dagger}$, where
		\begin{eqnarray}
		o_1^i&=&\sqrt{8/9} \sigma_x - 1/3 \sigma_z; \\
		o_2^i&=&-\sqrt{2/9} \sigma_x + \sqrt{2/3} \sigma_y - 1/3 \sigma_z; \nonumber\\
		o_3^i&=&-\sqrt{2/9} \sigma_x - \sqrt{2/3} \sigma_y - 1/3 \sigma_z;\nonumber\\
		o_4^i&=&\sigma_z. \nonumber
		\end{eqnarray}
with $i$ enumerates the observers ($i=1,2,3$) and $j$ - the measurement settings ($j=1,2,3,4$). 
For $10^7$ random settings we observe that the W state always exhibits genuine multipartite nonlocality violating the $\mathcal{NS}_2$ model ($P_v = 100\%$). 

Qualitatively, this result can be explained by the strong activation of the 47th facet inequality of $\mathcal{NS}_2$ polytope (hereafter $I_{47}^{\mathcal{NS}_2}$). For tetrahedrally distributed settings $I_{47}^{\mathcal{NS}_2}$ provides $\mathcal{P}_{V}^{I}=99.9997\%$ while uniform random measurements yield $\mathcal{P}_{V}^{I}=52.3103\%$. Moreover, our numerical research based on $n = 10^9$ random settings reveal that all those few measurements which did not provide violation of $I_{47}^{\mathcal{NS}_2}$ always give violation of $I_{4}^{\mathcal{NS}_2}$. Note that $\mathcal{P}_{V}^{I}$ for $I_{4}^{\mathcal{NS}_2}$ and $I_{16}^{\mathcal{NS}_2}$ with tetrahedrally distributed settings is equal to $99.998\%$ (c.f. Table \ref{tab_qubits}), so less than $I_{47}^{\mathcal{NS}_2}$.

On the other hand, if the distribution of $\mathcal{P}_{V}$ is tested against $\mathcal{S}_{max}$, then the inequality $I_{4}^{\mathcal{NS}_2}$ and $I_{16}^{\mathcal{NS}_2}$ gives a good approximation of the LP results for uniform random settings (see Fig. \ref{fig:tetrads}(b)). This result confirms the usefulness of $(I_{4}^{\mathcal{NS}_2},I_{16}^{\mathcal{NS}_2})$ for the study $\mathcal{NS}_2$-nonlocal correlations for W states. For tetrahedrally distributed settings the approximation of $\mathcal{P}_{V}$ to $\mathcal{S}_{max}$ is much worse for both $I_{47}^{\mathcal{NS}_2}$ and $(I_{4}^{\mathcal{NS}_2},I_{16}^{\mathcal{NS}_2})$. In other words, the inequalities $I_{47}^{\mathcal{NS}_2}$ and $I_{4}^{\mathcal{NS}_2}$ seem sufficient to explain the final value of $\mathcal{P}_{V}$ but not sufficient to explain details of $\mathcal{P}_{V}$ for tetrahedrally distributed settings.

Finally, it is worth noting that the restriction of measurement settings to those that form tetrahedrons impacts the distribution of nonlocality strange, shifting it toward higher values (Fig. \ref{fig:tetrads}(b)).

For the other states, settings with tetrahedral distribution also give an improvement of $\mathcal{P}_{V}$ over the uniform random measurements.
The probability of violation the model for the GHZ state is also high, $\mathcal{P}_{V} = 99.9986\%$, but less than 100\%. In this case, $I_{4}^{\mathcal{NS}_2}$ provides $\mathcal{P}_{V}^{I}=99.9777\%$ while $I_{47}^{\mathcal{NS}_2}$ provides $\mathcal{P}_{V}^{I}=0\%$.  For $\psi_s$, $\mathcal{P}_V = 99.9944 < 100\%$ and $\mathcal{P}^I=99.9693\%$. In both cases the statistics was $10^6$.
		
		

	\section{Three-qubit GHZ state is Svetlichny nonlocal for two orthogonal planar settings}
	
	We consider three observers that share the three qubit GHZ state and perform a Bell type experiment, in which they can choose between two orthogonal planar settings. The corresponding observables are given by $O_j^i = U^i o_j^i (U^i)^{\dagger}$, where $o_1^i = \sigma_x$, $o_2^i = \sigma_y$, $(i=1,2,3; j=1,2)$. In such scenario, for $10^7$ random set of settings, we observe that all probability points turn out to be genuine multipartite nonlocal. The computed critical visibilities are in the range: $0.707106781 \leq v_{crit} \leq 0.999999782 < 1$.
	
	 Restricting the measurements exclusively to the $x-y$ plane it can be analitically proved that measurements on the GHZ state are orthogonal for each 8 party in the $x-y$ plane and no aligned reference frames are needed to violate a Svetlichny-type inequality. There are analytical family of Svetlichny-type inequalities which detect GMNL for any such random orthogonal directions in the common plane $x-y$. The inequalities are in general not exactly the Svetlichny one, but they are related to it (we have to apply some appropriate rotation on the coefficients of the three-party Svetlichny inequality). 
	
    Let the observers choose the following observables:
\begin{eqnarray}
A_i &=& \cos\alpha_i \sigma_x + \sin\alpha_i \sigma_y,\nonumber\\
B_j &=& \cos\beta_j \sigma_x + \sin\beta_j \sigma_y,\nonumber\\
C_k &=& \cos\gamma_k \sigma_x + \sin\gamma_k \sigma_y,
\end{eqnarray}
where $\sigma_x$ and $\sigma_y$ are the Pauli matrices. These give the three-party correlations
\begin{equation}
\expect{A_iB_jC_k}=\bra{{\rm GHZ}}A_i\otimes B_j\otimes C_k\ket{{\rm GHZ}}=\cos(\alpha_i+\beta_j+\gamma_k)
\label{expABC}
\end{equation}
for $i,j,k=1,2$. In addition, the following measurement angles are chosen for the second setting:
\begin{align}
\alpha_2&=\alpha_1+\pi/2, \nonumber\\
\beta_2&=\beta_1+\pi/2, \nonumber\\
\gamma_2&=\gamma_1+\pi/2.
\label{angles2}
\end{align}
That is, the first and second settings are orthogonal to each other for each party. We will show that the quantum correlation~(\ref{expABC}) describing the experiment is genuinely tripartite nonlocal for generic angles $\alpha_1,\beta_1$ and $\gamma_1$.

To this end, we take the Svetlichny-type inequality
\begin{equation}
\sum_{(i,j,k)=1}^2S_{ijk}\langle A_iB_jC_k\rangle \le S_2,
\label{S2bound1}
\end{equation}
with the following coefficients:
\begin{equation}
S_{ijk}=\cos(\alpha_i+\beta_j+\gamma_k)
\label{Sijk}
\end{equation}
for $i,j,k=(1,2)$, where $\alpha_2,\beta_2$ and $\gamma_2$ are defined by ($\ref{angles2}$). Define the sum of the three angles $\alpha_1,\beta_1$ and $\gamma_1$:
\begin{equation}
\theta=\alpha_1+\beta_1+\gamma_1.
\label{theta}
\end{equation} 
Then, the Svetlichny inequality looks like this
\begin{equation}
\sum_{(i,j,k)=1}^2S_{ijk}\langle A_iB_jC_k\rangle \le 4\max{(|\cos\theta|,|\sin\theta|)},
\label{S2bound2}
\end{equation}
where the value on the right is the Svetlichny-type local bound, i.e. the bound that can be attained with $S_2$ local correlations. Note that the above inequality is a function of the three-parameters ($\alpha_1,\beta_1$ and $\gamma_1$).

{\it Quantum maximum with GHZ state}---First we calculate the quantum value of the Svetlichny expression with the GHZ state and the above measurement directions. This gives $4$ for any value of $\theta$. As we can see, this means that randomly generated measurement angles will always produce genuinely tripartite nonlocal Svetlichny correlations, apart from a set of zero measure. Using (\ref{angles2}), we obtain the following
\begin{align}
&S_{ij1}\expect{A_iB_jC_1}+S_{ij2}\expect{A_iB_jC_2}=\cos^2(\alpha_i+\beta_j+\gamma_1)\nonumber\\
&+\cos^2(\alpha_i+\beta_j+\gamma_1+\pi/2)=1
\end{align}
for all $i,j$. Therefore, summing up the above expression for all $(i,j)$ pairs, we have 
\begin{equation}
Q=\sum_{i,j=1}^2\left(S_{ij1}\langle A_iB_jC_1\rangle+S_{ij2}\langle A_iB_jC_2\rangle\right)=4.
\end{equation}

{\it Bilocal $S_2$ maximum}---We calculate the bound $S_2$ for bipartite nonlocality (i.e., we prove the right-hand side of (\ref{S2bound2})). It suffices to consider the $AB|C$ bipartition. The proof of the bound for the other two partitions $A|BC$ and $B|AC$ is analogous. For the partition $AB|C$, the correlation function is factorized as follows:
\begin{equation}
\expect{A_iB_jC_k}=\expect{A_iB_j}\expect{C_k}.
\end{equation}
Then $S_2$ is given by
\begin{equation}
S_2= \max\sum_{i,j,k}S_{ijk}\expect{A_iB_j}\expect{C_k},
\end{equation}
where the maximum is taken over all $\expect{C_k}$ and $\expect{A_iB_j}$ functions. Without loss of generality, to compute $S_2$, we can take the extremal values $\expect{C_k}=\pm 1$ for $k=1,2$ and $\expect{A_iB_j}=\pm 1$ for $i=1,2$, $j=1,2$. Then we have
\begin{equation}
S_2 = \max_{C_1=\pm1, C_2=\pm1}\sum_{i,j=1}^2{|C_1S_{ij1}+C_2S_{ij2}|}.
\end{equation}
Here we prove that 
\begin{equation}
\max_{C_1=\pm1, C_2=\pm1}\sum_{j=1}^2{|C_1S_{ij1}+C_2S_{ij2}|}\le 2\max{(|\cos\theta|,|\sin\theta|)}
\label{maxC12}
\end{equation}
for $i=1,2$. The value 2 on the right is therefore only achieved for special values of $\theta$. Indeed, if we explicitly write out $S_{ij1}$ and $S_{ij2}$ in (\ref{Sijk}) and denote 
\begin{equation}
\varphi_i=\alpha_i+\beta_1+\gamma_1,
\label{varphii}
\end{equation}
we get $\varphi_1=\theta$ and $\varphi_2=\theta+\pi/2$ by using (\ref{theta}), and the left-hand side of (\ref{maxC12}) will be
\begin{equation}
|C_1\cos\varphi_i+C_2\cos(\varphi_i+\pi/2)|+|C_1\cos(\varphi_i+\pi/2)-C_2\cos\varphi_i|.
\end{equation}

Now we use the simple relation $|a|+|b|=\max(|a+b|,|a-b|)$ for two real numbers $a$ and $b$ to get the bound
\begin{equation}
2\max(|\cos\varphi_i|,|\sin\varphi_i|)
\label{cossin}
\end{equation}
on (\ref{maxC12}), which in turn entails the bound $S_2\le 4\max{(|\cos\theta|,|\sin\theta|)}$. The value 4 is attained only if $\theta=k\pi/2$, where $k$ is an arbitrary integer. This implies $\alpha_1+\beta_1+\gamma_1=k\pi/2$. However, this is only a subset of measure zero of all possible triples ($\alpha_1,\beta_1,\gamma_1$). Therefore, in the case of generic $XY$ planar random orthogonal measurements on the GHZ state, the correlations will always be genuinely Svetlichny nonlocal.

\begin{figure}
		\centering
	\includegraphics[width=0.48\textwidth]{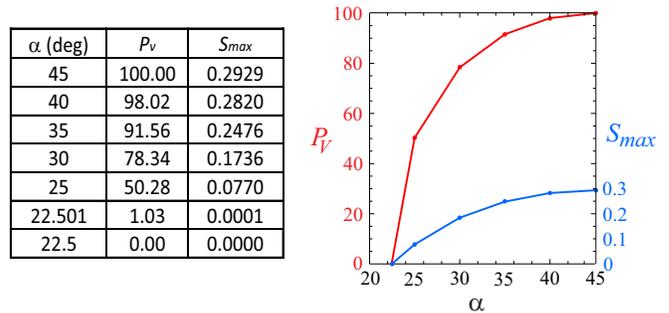}
		\caption{\label{GHZXY} Probability (red) and strength (blue) of violation of $\mathcal{NS}_2$ model by the asymmetric GHZ($\alpha$) state and measurement settings distributed on the x-y plane. }
		\end{figure}

It worth noting that for the generalized GHZ state $\cos \alpha |000\rangle + \sin \alpha |111\rangle$, if we break the symmetry and decrease the value of $\alpha$, we immediately start to observe cases that do not exhibit genuine multipartite nonlocality to finally make it disappear for $\alpha = 22.5^o$ (see Tab.~\ref{GHZXY}, and Fig.~\ref{GHZXY}). Note that the numerical results given in Tab.~\ref{GHZXY} are in agreement with $v_{crit}=1/[\sqrt{2}~\sin(2 \theta)]$, i.~e., the critical visibility for Svetlichny inequality.
	

\section{Conclusions}

We present a linear programming method for exploring the conflict of quantum predictions with local and realistic models for multipartite systems. Such conflicts lead to the conclusion that genuine multipartite nonlocality is present in our system. Moreover, the nature of our method implies its optimality. We show some examples for systems of three particles (qubits and qutrits). By introducing some restrictions on the distribution of observables, we obtain guaranteed nonlocality for the W and GHZ states. In the first case, we sample four observables such that their corresponding direction vectors form a tetrahedron. In the second case, we deal with two observables lying in the same (x-y) plane.  

\section*{Acknowledgments}

M.P. was partially supported by National Science Centre (Poland) grants: 2016/23/G/ST2/04273 and 2017/26/E/ST2/01008. A.B. acknowledges financial support by the Czech Science Foundation under the project No.~20-17765S and  project CZ.02.1.01/0.0/0.0/16\_019/0000754 of the Ministry of Education, Youth and Sports of the Czech Republic. Numerical calculations were performed in the Wroclaw Centre for Networking and Supercomputing, Poland. M.I. and T.V. acknowledge the support of the EU (QuantERA eDICT) and the National Research, Development and Innovation Office NKFIH (No. 2019-2.1.7-ERA-NET-2020-00003). W.L. acknowledges support by the Foundation for Polish Science (IRAP project ICTQT, Contract No. 2018/MAB/5, cofinanced by EU via Smart Growth Operational Programme).

\end{document}